\documentclass[aps,pra,twocolumn,superscriptaddress,showpacs]{revtex4-1}
\usepackage[version=3]{mhchem}
\usepackage[english]{babel}
\usepackage{dcolumn}
\usepackage{graphicx}
\usepackage{amsmath}
\usepackage{longtable}
\usepackage{bigstrut}
\usepackage[colorlinks=true, urlcolor=blue, citecolor=blue, filecolor=blue, linkcolor=blue]{hyperref}

\begin{document}

\title{Methyl mercaptan (\ce{CH3SH}) as a probe for $\mu$ variation}

\author{Paul Jansen}
\affiliation{Institute for Lasers, Life and Biophotonics, VU University Amsterdam, 
De Boelelaan 1081, 1081 HV Amsterdam, The Netherlands}
\author{Li-Hong Xu}
\affiliation{Department of Physics and Centre for Laser, Atomic, and Molecular Sciences, University of New Brunswick, Saint John, New Brunswick E2L 4L5, Canada}
\author{Isabelle Kleiner}
\affiliation{Laboratoire Interuniversitaire des Syst\`{e}mes Atmosph\'{e}riques (LISA), CNRS UMR 7583 et Universit\'{e}s Paris Diderot et Paris Est, 61 av. G\'{e}n\'{e}ral de Gaulle, 94010 Cr\'{e}teil C\'{e}dex, France}
\author{Hendrick L. Bethlem}
\affiliation{Institute for Lasers, Life and Biophotonics, VU University Amsterdam, 
De Boelelaan 1081, 1081 HV Amsterdam, The Netherlands}
\author{Wim Ubachs}
\affiliation{Institute for Lasers, Life and Biophotonics, VU University Amsterdam, 
De Boelelaan 1081, 1081 HV Amsterdam, The Netherlands}

\date{\today}

\begin{abstract}
Torsion-rotation transitions in molecules exhibiting hindered internal rotation possess enhanced sensitivities to a variation of the proton-to-electron mass ratio. This enhancement occurs due to a cancellation of energies associated with the torsional and rotational degrees of freedom of the molecule. This effect occurs generally in every internal rotor molecule, but is exceptionally large in methanol. In this paper we calculate the sensitivity coefficients of methyl mercaptan, the thiol analogue of methanol. The obtained sensitivity coefficients in this molecule range from $K_\mu=-14.8$ to $+12.2$ for transitions with a lower-level excitation energy below 10\,cm$^{-1}$.
\end{abstract}


\maketitle
\section{Introduction}
Physical theories extending the Standard Model of particle physics have presented scenarios that allow for spatial-temporal variations of the constants of nature~\cite{Uzan2003}. Since the initial findings of a possible variation of the fine structure constant by Webb~\emph{et al.}~\cite{Webb1999} there has arisen a great activity in search for signatures of such variations. Studies aimed at detecting a possible drift of a fundamental constant on a cosmological time scale focus mainly on  the fine structure constant, $\alpha$,~\cite{Webb2001,Murphy2003,Webb2011,King2012} and the proton-to-electron mass ratio, $\mu$~\cite{Reinhold2006,Malec2010,vanWeerdenburg2011}. A variation of $\alpha$ or $\mu$ will manifest itself as a change in the spectrum of atoms and molecules, since such a variation may induce a shift in the position of a spectral line. Not all lines will shift in the same amount or direction. The response of a transition to a variation of $\alpha$ or $\mu$ is characterized by its sensitivity coefficient, $K_\mu$ or $K_\alpha$, respectively, which is defined as the proportionality constant between the fractional frequency shift of the transition, $\Delta\nu/\nu$ and the fractional shift in $\alpha$ or $\mu$. 

\begin{equation}
\frac{\Delta\nu}{\nu}=K_X\frac{\Delta X}{X},\quad\text{with } X = \alpha,\mu.
\label{eq:Kmu}
\end{equation}

The search for $\mu$ variation on a cosmological time scale has been made operational by comparing optical transitions of molecular hydrogen (\ce{H2}) in high-redshifted objects with accurate laboratory measurements~\cite{Reinhold2006}. These investigations have yielded a limit at the level of  $\Delta\mu/\mu < 10^{-5}$ for look-back times of 12 billion years~\cite{Malec2010,vanWeerdenburg2011}. The transitions in \ce{H2} that were used to obtain this result possess sensitivity coefficients that range from $-0.05$ to $+0.02$. Inversion transitions of ammonia (\ce{NH3}) were found to be $\sim$100 times more sensitive to $\mu$-variation than \ce{H2} transitions~\cite{Veldhoven2004,Flambaum&Kozlov2007}. Astronomical observations of \ce{NH3}, in the microwave or radio range of the electromagnetic spectrum, led to stringent 1$\sigma$ constraints at the level of $(1.0 \pm 4.7) \times 10^{-7}$~\cite{Murphy2008} and $(–3.5 \pm 1.2) \times 10^{-7}$~\cite{Henkel2009}. Soon thereafter it was realized that the large number of degrees of freedom that exist in even the simplest polyatomic molecules can result in large enhancements of the sensitivity coefficients for a possible drift in $\mu$. These enhancements occur for transitions between near-degenerate levels that each have a different dependence on $\mu$. For instance, it was found that mixed inversion-rotation transitions in \ce{H3O+} have sensitivity coefficients ranging from $K_\mu=-9$ to $+5.7$~\cite{Kozlov2011}, while the Renner-Teller interaction in {\it l}-\ce{CH3} results in sensitivity coefficients ranging from $K_\mu=-53$ to $+742$~\cite{Kozlov2013}. Mixed torsion-wagging-rotation transitions in methylamine display sensitivity coefficients ranging from $K_\mu=-19.1$ to $-0.75$~\cite{Ilyushin2012}. 

In the context of astrophysical searches, methanol~\cite{Jansen2011PRL,Levshakov2011} is the target species of choice since it possesses sensitive transitions at low excitation energy and has been observed at high redshift~\cite{Muller2011}. In a recent study, Bagdonaite~\emph{et al.} used four transitions in methanol to constrain $\Delta\mu/\mu$ at $(0.0\pm 1.0)\times 10^{-7}$ at a look-back time of 7 billion years~\cite{Bagdonaite2013}. Methyl mercaptan (\ce{CH3SH}) is the sulfur analogue of methanol and might therefore posses transitions that have large sensitivity coefficients to a variation of $\mu$~\cite{Jansen2011}. Although methyl mercaptan has thus far only been detected in our local galaxy~\cite{Linke1979,Gibb2000}, recent advances in radio telescopes have greatly increased the number of detected molecular species at high redshift. It is therefore relevant to have a list available that contains the sensitivity coefficients of transitions in methyl mercaptan that might be observed in the interstellar medium. 

The recent terahertz and far-infrared study of the normal isotopologue (\ce{^{12}CH3^{32}SH}) of methyl mercaptan by Xu~\emph{et al.}~\cite{Xu2012} has resulted in a complete list of the molecular parameters for this molecule. In this paper we use the results of Xu~\emph{et al.}~\cite{Xu2012} and the scaling relations of the molecular parameters that were derived in Jansen~\emph{et al.}~\cite{Jansen2011PRL,Jansen2011} to calculate the sensitivity coefficients of methyl mercaptan. 

\section{Structure of methyl mercaptan}
\begin{figure}[tb]
\includegraphics[width=\columnwidth]{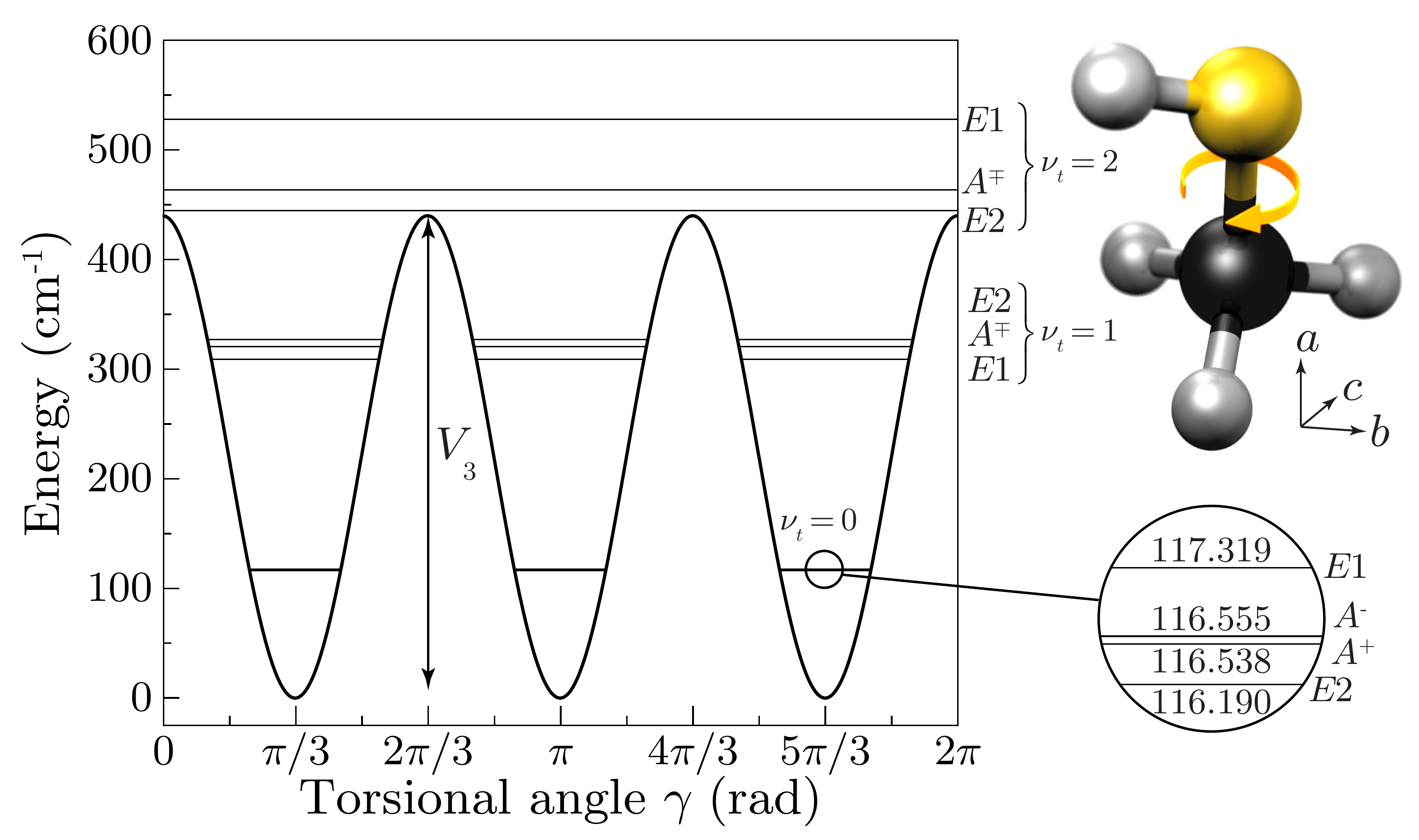}
\caption{\emph{(color online)} Variation of the potential energy of methyl mercaptan with the relative rotation $\gamma$ of the \ce{SH}-group with respect to the methyl group about the molecular axis. Shown are the $J = 1, |K| = 1$ energies of the lowest torsion-vibrational levels. The splitting between the different symmetry levels is due to tunneling through the potential barriers. The $A$-symmetry species are split further due to the asymmetry of the molecule ($K$-splitting). 
\label{fig:potential_and_structure}}
\end{figure}

Methyl mercaptan or methanethiol, depicted on the right-hand side of Fig.~\ref{fig:potential_and_structure}, consists of a thiol group (\ce{SH}) attached to a methyl group (\ce{CH3}) and is thus the sulfur analogue of methanol. The \ce{CS} sigma bond connecting the two parts of the molecule is flexible, allowing the methyl group to rotate with respect to the thiol group. As in the case of methanol, this rotation is not free but hindered by a threefold potential barrier with minima and maxima that correspond to the staggered and eclipsed configuration of the molecule, respectively. For the lowest energy levels this relative or internal rotation is classically forbidden and only occurs due to quantum mechanical tunneling of the hydrogen atoms. As a consequence of this tunneling each rotational level splits into three levels that are labeled according to their symmetry as $A$ or $E$, as can be seen in on the left-hand side of Fig.~\ref{fig:potential_and_structure}.

\begin{figure*}[tb]
\includegraphics[width=0.8\textwidth]{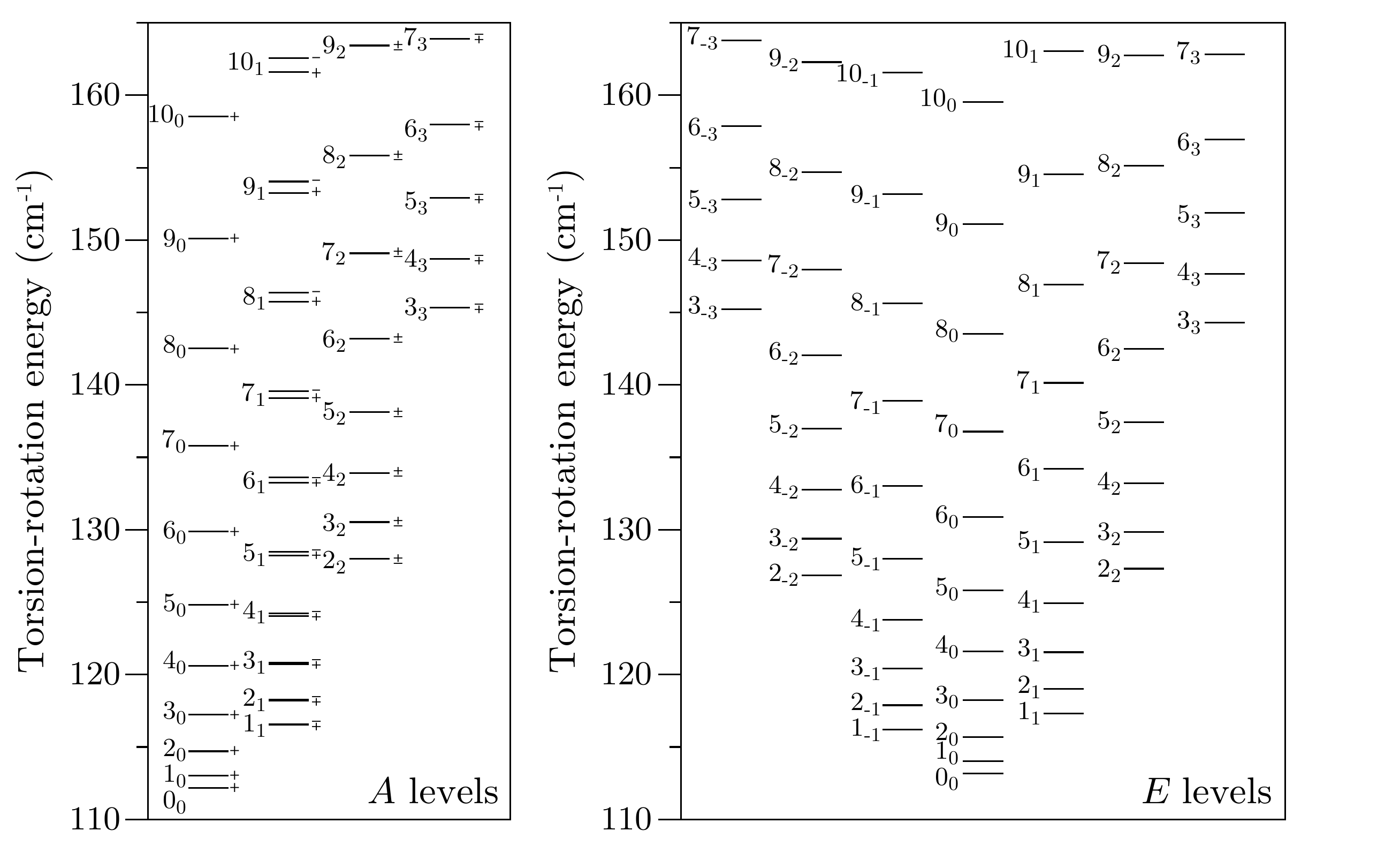}
\caption{Energy of the lowest rotational levels in the torsion-vibrational ground state ($\nu_{t} = 0$) of methyl mercaptan \ce{(^{12}CH3^{32}SH)}~\cite{Xu2012}. The levels are denoted by $J_K$ (indicated on the left side of each level). For the $A$ levels the so-called parity quantum number ($+/-$) is also used. The panel on the left displays the $A$ state levels, whereas the panel on the right displays the $E$ state levels. High sensitivities are expected for transitions that connect near degenerate levels with different $K$. 
\label{fig:K-ladders}}
\end{figure*}

The lowest energy levels of \ce{CH3SH} are shown in the left and right panel of Fig.~\ref{fig:K-ladders}. The $A$ and $E$ species can be considered as two different molecular species in the same sense as ortho- and para-ammonia, respectively. The arrangement of energy levels within a symmetry state resembles that of a prolate symmetric top, with the difference being that every $K$ ladder obtains a small energy offset due to the $K$ dependent tunneling splitting. 

As a consequence, certain states in neighboring $K$ ladders may become near degenerate which results in a large enhancement of the sensitivity coefficients $K_\mu$ for transitions connecting these states~\cite{Jansen2011PRL,Levshakov2011}.

\section{Sensitivity coefficients}
The energy levels of methyl mercaptan have been calculated using an adapted version of the {\sc belgi} code~\cite{Hougen1994}. This code was modified and improved by Xu \emph{et al.}~\cite{Xu2008} in a number of ways useful for treating the large datasets available for the methanol molecule, but the general approach has not been changed compared to the initial code. The present version of the code has been sped up compared to the original version, and also a substantial number of higher order parameters has been added. Using the set of 79 molecular parameters of methyl mercaptan obtained by Xu~\emph{et al.}, the lower energy levels are found with an accuracy $<100$\,kHz. 

In order to calculate the $K_{\mu}$ coefficients for the different transitions in methyl mercaptan, the energy of each level and its dependence on $\mu$ has to be obtained. This translates into knowing the values of the molecular constants that go into {\sc belgi} and how these constants scale with $\mu$. These scaling relations were obtained in a previous study~\cite{Jansen2011PRL}, while a more detailed discussion on the derivation can be found in Ref.~\cite{Jansen2011}. 

Table~\ref{tab:sensitivities} lists calculated transition frequencies and sensitivity coefficients in methyl mercaptan with an excitation energy less than 10\,cm$^{-1}$, that is, those transitions most relevant for astrophysical searches. It can be seen that several of these transitions display relatively large sensitivity coefficients. 

\begingroup
\squeezetable
\renewcommand{\arraystretch}{1.1}
\begin{table*}[bt]
\caption{Transition frequencies and sensitivity coefficients in methyl mercaptan with a lower-level excitation energy of less than 10\,cm$^{-1}$, calculated with {\sc belgi} using the molecular constants of Ref.~\cite{Xu2012} (fifth column) and the toy model of Ref.~\cite{Jansen2011} (sixth column). The third column lists the transition strength multiplied by the electric dipole moment, $\mu_{e}$, squared, while the fourth column lists the excitation energy of the lower level in Kelvin. 
\label{tab:sensitivities}}
\begin{tabular}{ll}
\begin{tabular}[t]{>{$}l<{$} D..{3} D..{3} D..{1} D..{2} D..{2}}
\toprule
\multicolumn{1}{l}{Transition} & 
\multicolumn{1}{l}{Energy (MHz)} &
\multicolumn{1}{l}{$S\mu_e^2$ (D$^2$)} &
\multicolumn{1}{l}{$T_\text{low}$ (K)} &
\multicolumn{1}{l}{$K_\mu^{\text{\sc belgi}}$} &
\multicolumn{1}{l}{$K_\mu^\text{toy}$}\\
\colrule

1_{1} \rightarrow	1_{1}	A^\pm	& 523.147	& 1.219 & 6.3 & -0.98 & -1.00 \\
2_{1}	\rightarrow	2_{1}	A^\pm	& 1569.410	& 0.677 & 8.7 & -0.98 & -1.00 \\
3_{1}	\rightarrow	4_{0}	E		& 1874.635	& 1.390 & 12.0 & 12.20 & 11.77 \\
4_{0}	\rightarrow	3_{1}	A^+	& 3038.566	& 3.077 & 12.1 & -14.83 & -14.94 \\
3_{1}	\rightarrow	3_{1}	A^\pm	& 3138.723	& 0.473 & 12.3 & -0.98 & -1.00 \\
2_{-1}\rightarrow	3_{0}	E	& 10534.181	& 1.064 & 6.8 & -7.55 & -7.29 \\
2_{0}	\rightarrow	1_{-1}	E	& 14764.687	& 0.513 & 3.6 & 3.68 & 3.49 \\
3_{0}	\rightarrow	2_{1}	E		& 23339.083	& 0.966 & 7.3 & -2.07 & -2.03 \\
0_{0}	\rightarrow	1_{0}	A^+	& 25290.869	& 0.813 & 0.0 & -1.00 & -1.00 \\
0_{0}	\rightarrow	1_{0}	E		& 25291.824	& 0.814 & 0.0 & -1.00 & -1.00 \\
3_{0}	\rightarrow	2_{1}	A^+	& 29091.802	& 2.038 & 7.3 & -2.44 & -2.46 \\
3_{-1}\rightarrow	4_{0}	E		& 35857.370	& 1.678 & 10.4 & -2.92 & -2.85 \\
2_{0}	\rightarrow	1_{1}	E		& 48604.208	& 0.496 & 3.6 & -1.51 & -1.49 \\
1_{1}	\rightarrow	2_{1}	A^+	& 50058.794	& 1.220 & 6.3 & -1.00 & -1.00 \\
1_{-1}\rightarrow	2_{-1}	E	& 50565.538	& 1.220 & 4.3 & -1.00 & -1.00 \\
1_{0}	\rightarrow	2_{0}	A^+	& 50579.301	& 1.625 & 1.2 & -1.00 & -1.00 \\
1_{0}	\rightarrow	2_{0}	E		& 50580.882	& 1.629 & 1.2 & -1.00 & -1.00 \\
1_{1}	\rightarrow	2_{1}	E		& 50599.280	& 1.221 & 6.0 & -1.00 & -1.00 \\
1_{1}	\rightarrow	2_{1}	A^-	& 51105.057	& 1.220 & 6.3 & -1.00 & -1.00 \\
2_{0}	\rightarrow	1_{1}	A^+	& 54895.867	& 1.014 & 3.6 & -1.76 & -1.77 \\
4_{0}	\rightarrow	4_{-1}	E	& 65172.338	& 3.774 & 12.1 & 0.05 & 0.02 \\
3_{0}	\rightarrow	3_{-1}	E	& 65282.263	& 3.162 & 7.3 & 0.06 & 0.02 \\
2_{0}	\rightarrow	2_{-1}	E	& 65330.225	& 2.383 & 3.6 & 0.06 & 0.01 \\
1_{0}	\rightarrow	1_{-1}	E	& 65345.568	& 1.480 & 1.2 & 0.06 & 0.01 \\
2_{1}	\rightarrow 3_{1}	A^+	& 75085.877	& 2.168 & 8.7 & -1.00 & -1.00 \\
2_{-1}\rightarrow	3_{-1}	E	& 75816.443	& 2.168 & 6.8 & -1.00 & -1.00 \\
2_{0}	\rightarrow	3_{0}	A^+	& 75862.860	& 2.438 & 3.6 & -1.00 & -1.00 \\
2_{0}	\rightarrow	3_{0}	E	   & 75864.406	& 2.443 & 3.6 & -1.00 & -1.00 \\
2_{1}	\rightarrow	3_{1}	E	   & 75925.915	& 2.169 & 8.4 & -1.00 & -1.00 \\
2_{1}	\rightarrow	3_{1}	A^-	& 76655.189	& 2.168 & 8.8 & -1.00 & -1.00 \\
0_{0}	\rightarrow	1_{-1}	E	& 90637.393	& 1.017 & 0.0 & -0.24 & -0.27 \\
1_{0}	\rightarrow	1_{1}	E	   & 99185.090	& 1.520 & 1.2 & -1.25 & -1.24 \\
2_{0}	\rightarrow	2_{1}	E	   & 99203.488	& 2.602 & 3.6 & -1.25 & -1.24 \\
3_{0}	\rightarrow	3_{1}	E	   & 99264.998	& 3.784 & 7.3 & -1.25 & -1.24 \\
4_{0}	\rightarrow	4_{1}	E	   & 99409.714	& 5.101 & 12.1 & -1.24 & -1.24 \\
3_{1}	\rightarrow 4_{1}	A^+	& 100110.190 & 3.049 & 12.3 & -1.00 & -1.00 \\
3_{-1}\rightarrow	4_{-1}	E	& 101029.708 &3.048 & 10.4 & -1.00 & -1.00 \\
3_{0}	\rightarrow	4_{0}	A^+	& 101139.112 & 3.251 & 7.3 & -1.00 & -1.00 \\
3_{0}	\rightarrow	4_{0}	E	   & 101139.633 & 3.257 & 7.3 & -1.00 & -1.00 \\
3_{1}	\rightarrow	4_{1}	E	   & 101284.349 & 3.049 & 12.0 & -1.00 & -1.00 \\
3_{1}	\rightarrow	4_{1}	A^-	& 102202.438 & 3.049 & 12.4 & -1.00 & -1.00 \\
1_{0}	\rightarrow	1_{1}	A^\pm	& 105998.315 & 3.018 & 1.2 & -1.40 & -1.40 \\
2_{0}	\rightarrow	2_{1}	A^\pm	& 106524.072 & 5.017 & 3.6 & -1.39 & -1.40 \\

\botrule
\end{tabular}&


\begin{tabular}[t]{>{$}l<{$} D..{3} D..{3} D..{1} D..{2} D..{2}}
\toprule
\multicolumn{1}{l}{Transition} & 
\multicolumn{1}{l}{Energy (MHz)} &
\multicolumn{1}{l}{$S\mu_e^2$ (D$^2$)} &
\multicolumn{1}{l}{$T_\text{low}$ (K)} &
\multicolumn{1}{l}{$K_\mu^{\text{\sc belgi}}$} &
\multicolumn{1}{l}{$K_\mu^\text{toy}$}\\
\colrule

3_{0}	\rightarrow	3_{1}	A^\pm	& 107316.401 & 6.996 & 7.3 & -1.39 & -1.39 \\
4_{0}	\rightarrow	4_{1}	A^\pm	& 108379.727 & 8.947 & 12.1 & -1.39 & -1.39 \\
1_{0}	\rightarrow	2_{-1}	E	& 115911.107 & 1.572 & 1.2 & -0.40 & -0.43 \\ 
0_{0}	\rightarrow	1_{1}	E	   & 124476.914 & 0.983 & 0.0 & -1.20 & -1.19 \\
0_{0}	\rightarrow	1_{1}	E	   & 124476.914 & 0.983 & 0.0 & -1.20 & -1.19 \\
4_{0}	\rightarrow	5_{0}	E	   & 126403.807 & 4.071 & 12.1 & -1.00 & -1.00 \\
4_{0}	\rightarrow	5_{0}	A^+	& 126405.629 & 4.063 & 12.1 & -1.00 & -1.00 \\
0_{0}	\rightarrow	1_{1}	A^+	& 130766.037 & 2.012 & 0.0 & -1.32 & -1.32 \\
2_{0}	\rightarrow	3_{-1}	E	& 141146.668 & 2.190 & 3.6 & -0.51 & -0.53 \\
1_{0}	\rightarrow	2_{1}	E	   & 149784.370 & 1.428 & 1.2 & -1.17 & -1.16 \\
1_{0}	\rightarrow	2_{1}	A^+	& 155533.962 & 3.018 & 1.2 & -1.27 & -1.27 \\
3_{0}	\rightarrow	4_{-1}	E	& 166311.971 & 2.896 & 7.3 & -0.59 & -0.60 \\
3_{1}	\rightarrow	2_{2}	E	   & 172960.517 & 0.333 & 12.0 & -0.64 & -0.67 \\
2_{0}	\rightarrow	3_{1}	E	   & 175129.404 & 1.816 & 3.6 & -1.14 & -1.14 \\
2_{0}	\rightarrow	3_{1}	A^+	& 180040.538 & 4.029 & 3.6 & -1.23 & -1.24 \\
4_{0}	\rightarrow	5_{-1}	E	& 191366.909 & 3.710 & 12.1 & -0.64 & -0.65 \\ 
3_{-1}\rightarrow	2_{-2}	E	& 193330.838 & 0.335 & 10.4 & -0.97 & -0.98 \\
3_{0}	\rightarrow	4_{1}	E	   & 200549.347 & 2.127 & 7.3 & -1.12 & -1.12 \\
3_{0}	\rightarrow	4_{1}	A^+	& 204287.868 & 5.050 & 7.3 & -1.21 & -1.21 \\
3_{1}	\rightarrow	2_{2}	A^-	& 215287.322 & 0.337 & 12.4 & -1.28 & -1.28 \\
3_{1}	\rightarrow	2_{2}	A^+	& 218428.095 & 0.332 & 12.3 & -1.27 & -1.27 \\
4_{0}	\rightarrow	5_{1}	E	   & 226093.110 & 2.345 & 12.1 & -1.10 & -1.11 \\
4_{0}	\rightarrow	5_{1}	A^+	& 228279.583 & 6.084 & 12.1 & -1.18 & -1.19 \\
3_{1}	\rightarrow	3_{2}	E	   & 248835.455 & 2.929 & 12.0 & -0.75 & -0.77 \\
2_{1}	\rightarrow	2_{2}	E	   & 248886.432 & 1.672 & 8.4 & -0.75 & -0.77 \\
2_{-1}\rightarrow	2_{-2}	E	& 269147.282 & 1.684 & 6.8 & -0.98 & -0.99 \\
3_{-1}\rightarrow	3_{-2}	E	& 269204.583 & 2.949 & 10.4 & -0.98 & -0.99 \\
3_{1}	\rightarrow	3_{2}	A^\mp	& 291169.820 & 2.957 & 12.4 & -1.20 & -1.20 \\
2_{1}	\rightarrow	2_{2}	A^\mp	& 291944.561 & 1.680 & 8.8 & -1.20 & -1.20 \\
2_{1}	\rightarrow	2_{2}	A^\pm	& 293511.922 & 1.667 & 8.7 & -1.20 & -1.20 \\
3_{1}	\rightarrow	3_{2}	A^\pm	& 294298.297 & 2.911 & 12.3 & -1.20 & -1.20 \\
1_{1}	\rightarrow	2_{2}	E	   & 299485.712 & 2.994 & 6.0 & -0.79 & -0.81 \\
1_{-1}\rightarrow	2_{-2}	E	& 319712.820 & 3.015 & 4.3 & -0.98 & -0.99 \\
2_{1}	\rightarrow	3_{2}	E	   & 324761.370 & 3.315 & 8.4 & -0.81 & -0.83 \\
1_{1}	\rightarrow	2_{2}	A^-	& 343047.569 & 3.001 & 6.3 & -1.17 & -1.17 \\
1_{1}	\rightarrow	2_{2}	A^+	& 343572.765 & 2.993 & 6.3 & -1.17 & -1.17 \\
2_{-1}\rightarrow	3_{-2}	E	& 345021.026 & 3.337 & 6.8 & -0.98 & -0.99 \\
3_{1}	\rightarrow	4_{2}	E	   & 350003.733 & 3.711 & 12.0 & -0.82 & -0.84 \\
2_{1}	\rightarrow	3_{2}	A^-	& 367814.764 & 3.334 & 8.8 & -1.16 & -1.16 \\
2_{1}	\rightarrow	3_{2}	A^+	& 369394.420 & 3.308 & 8.7 & -1.16 & -1.16 \\
3_{-1}\rightarrow	4_{-2}	E	& 370371.709 & 3.732 & 10.4 & -0.98 & -0.99 \\
3_{1}	\rightarrow	4_{2}	A^-	& 392318.890 & 3.754 & 12.4 & -1.15 & -1.15 \\
3_{1}	\rightarrow	4_{2}	A^+	& 395488.347 & 3.695 & 12.3 & -1.15 & -1.15 \\

\botrule
\end{tabular}\\
\end{tabular}
\end{table*}
\renewcommand{\arraystretch}{1.1}
\endgroup

In the last column of Table~\ref{tab:sensitivities} the results of the approximate model of Ref.~\cite{Jansen2011} are listed. This ``toy" model is derived for molecules that exhibit hindered internal rotation and contain a $C_{3v}$ symmetry group. The model decomposes the energy of the molecule into a pure rotational and a pure torsional part. The rotational part is approximated by the well-known expression for the rotational energy levels of a slightly asymmetric top

\begin{equation}
E_\text{rot}(J,K)=\frac{1}{2}\left (B+C \right )J\left (J+1 \right )+\left (A-\frac{B+C}{2} \right )K^2,
\end{equation}

\noindent
with $A=3.428$\,cm$^{-1}$, $B=0.432$\,cm$^{-1}$, and $C=0.413$\,cm$^{-1}$ the rotational constants along the $a$, $b$, and $c$ axis of the molecule, respectively. The torsional energy contribution is approximated by a Fourier expansion as

\begin{equation}
E_\text{tors}(K)=F \left [a_0+a_1\cos\left \{ \frac{2\pi}{3}\left (\rho K +\sigma \right )\right \} \right ],
\label{eq:Etors}
\end{equation}

\noindent
with $F=15.040$\,cm$^{-1}$ the constant of the internal rotation, $\rho=0.652$ a dimensionless constant reflecting the coupling between internal and overall rotation, and $\sigma=0,\pm 1$ a constant relating to the torsional symmetry. The expansion coefficients $a_0$ and $a_1$ depend on the shape of the torsional potential. Since we are mainly interested in the torsional energy difference $a_0$ cancels, $a_1$ follows from

\begin{figure}[tb]
\includegraphics[width=0.95\columnwidth]{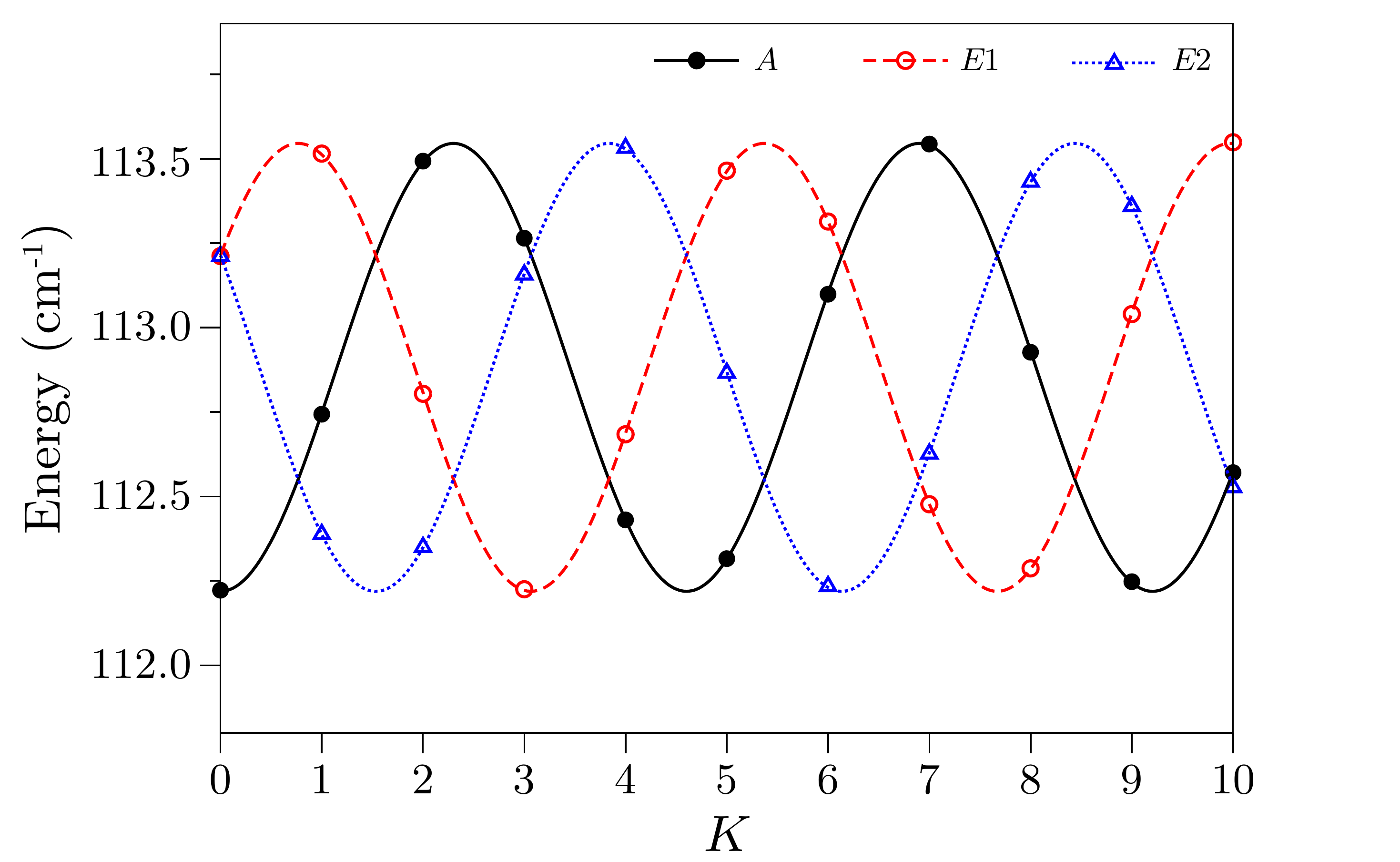}
\caption{\emph{(color online)} Torsional energies in the ground torsional state ($\nu_t=0$) of methyl mercaptan obtained with {\sc belgi} for $A$ (solid circles), $E1$ (open circles), and $E2$ (open triangles) levels as function of $K$. The solid, dashed, and dotted curves are fits to Eq.~\eqref{eq:Etors} for $A$, $E1$, and $E2$ states. Note that only integer values of $K$ have physical meaning. 
\label{fig:torsional_energies}}
\end{figure} 

\begin{equation}
a_1 = A_1s^{B_1}e^{-C_1\sqrt{s}},
\end{equation}

\noindent
with $A_1=-5.296$, $B_1=1.111$, and $C_1=2.120$~\cite{Jansen2011}. The dimensionless parameter $s=4V_3/9F$, with $V_3=441.442$\,cm$^{-1}$ the height of the barrier, is a measure of the effective potential. The torsional energy for methyl mercaptan is plotted in Fig.~\ref{fig:torsional_energies} as a function of $K$. Note that the torsional splitting between the $A$ and $E$ levels in the $K=0$ state of methyl mercaptan is 0.99\,cm$^{-1}$, and thus an order of magnitude smaller than in methanol, which has a torsional splitting of 9.1\,cm$^{-1}$. As a consequence, the amount of energy that can be cancelled in methyl mercaptan will be less than in methanol. 

Finally, the sensitivity coefficient of the transition is obtained from
\begin{equation}
K_\mu^\text{toy}=\frac{K_\mu^\text{rot}\Delta E_\text{rot}+K_\mu^\text{tors}\Delta E_\text{tors}}{\Delta E_\text{rot}+\Delta E_\text{tors}}.
\end{equation}

\noindent
Note that rather than using $\Delta{E_{\text{rot}}}+ \Delta{E_{\text{tors}}}$, we chose to use the experimental energy difference between the levels, $h \nu$, in order to account for the slight asymmetry of the molecule. The sensitivity of a pure rotational transition is $K_\mu^\text{rot}=-1$, whereas the sensitivity of a pure torsional transition is given by $K_\mu^\text{tors}=(B_1-1)-\tfrac{1}{2}C_1\sqrt{s}=-3.7$~\cite{Jansen2011}. The sensitivity coefficients of this simple model are seen to agree well with the results obtained by a diagonalization of the full molecular Hamiltonian, which reflects the robustness of the obtained results.

\section{Conclusions}
In this paper we have calculated sensitivity coefficients for transitions between low lying rotation levels in methyl mercaptan. The reported sensitivities span a range from $K_\mu=-14.8$ to $+12.2$ and can therefore be used to search for variation of $\mu$ in methyl mercaptan only. Although, thus far methyl mercaptan has only been detected in our local galaxy, it is our hope that the advanced spectral coverage, resolution, and sensitivity of the new generation of radio telescopes such as ALMA (Atacama Large Millimeter Array) will result in the detection of this molecule at high redshift. Note that the comprehensive line list of accurate rest frequencies for methyl mercaptan obtained by Xu~\emph{et al.}~\cite{Xu2012} should also alleviate this search. 

\section*{Acknowledgements}
RB acknowledges financial support from NWO via a VIDI-grant and by the ERC via a Starting Grant. LHX gratefully acknowledges financial support from the Natural Science and Engineering Research Council of Canada. IK acknowledges the financial support provided by ANR-08-BLAN-0054. PJ and WU acknowledge financial support from the Templeton Foundation.

\end{document}